\documentclass[%
 aip,%
 amsmath,amssymb,
reprint,%
]{revtex4-1}

\usepackage{graphicx}
\usepackage{dcolumn}
\usepackage{bm}
\usepackage{epstopdf}
\usepackage{booktabs}
\usepackage{multirow}
\usepackage{array}
\newcolumntype{P}[1]{>{\centering\arraybackslash}p{#1}}

\begin{document}


\title{Studies of non-trivial band topology and electron-hole compensation in YSb }

\author{Payal Wadhwa}
 \affiliation{T-GraMS Laboratory, Department of Physics, Indian Institute of Technology Ropar, Rupnagar, Punjab - 140001, India}
\author{Shailesh Kumar}%
\affiliation{School of Chemistry, Physics and Mechanical Engineering, Queensland University of Technology, Brisbane, Queensland 4000,\ Australia}
\affiliation{Manufacturing Flagship, CSIRO, Lindfield West, New South Wales 2070, Australia}

\author{Alok Shukla}
\affiliation{Department of Physics, Indian Institute of Technology Bombay, Powai, Mumbai - 400076,\ India}

\author{Rakesh Kumar}
\email{rakesh@iitrpr.ac.in}
\affiliation{T-GraMS Laboratory, Department of Physics, Indian Institute of Technology Ropar, Rupnagar, Punjab - 140001, India}

\date{\today}

\begin{abstract}
In this article, we study non-trivial topological phase and electron-hole compensation in extremely large magnetoresistance (XMR) material YSb under hydrostatic pressure using first-principles calculations.\ YSb is topologically trivial at ambient pressure, but undergoes a reentrant topological phase transition under hydrostatic pressure.\ The reentrant behavior of topological quantum phase is then studied as a function of charge density ratio under pressure.\ From the detailed investigation of Fermi surfaces, it is found that electron to hole densities ratio increases with pressure, however a non-trivial topological phase appears without perfect electron-hole compensation.\ The results indicate that the non-trivial topological phase under hydrostatic pressure may not have maximal influence on the magnetoresistance, and need further investigations through experiments to determine the exact relationship between topology and XMR effect.

\end{abstract}

\maketitle
\section{\label{sec:level1}Introduction}

Topological insulators (TI) are of supreme interest to the scientific community in recent years because of their extraordinary properties for applications in quantum computing and spintronics \cite{RevModPhys.82.3045,RevModPhys.83.1057,tian2018topological,da2018topological,adhip}.\ With protection by time-reversal symmetry,\ topological insulators possess intriguing physical properties like gapless surface states and unconventional spin texture with forbidden electron's backscattering \cite{RevModPhys.82.3045,PhysRevLett.95.146802,bernevig2006quantum}.\ More interestingly,\ topological phases of matter made an important breakthrough in physics theory,\ as not being characterized by symmetry breaking process like the one in conventional phase transitions given by Landau \cite{PhysRevLett.49.405}.\ Recently,\ the research on non-trivial band topology has been directed to semi-metals \cite{burkov2016topological,barik2018multiple,kong2018topological,PhysRevLett.121.226401,PhysRevLett.121.086804}, which could establish many phenomena such as quantum magnetoresistance \cite{li2016negative},\ chiral anomaly \cite{PhysRevX.5.031023},\ and Weyl fermion quantum transport \cite{vazifeh2013electromagnetic}.\\

More recently,\ extremely large magnetoresistance (XMR) materials like WTe$_{2}$ \cite{ali2014large,jiang2015signature}, Bi$_{2}$Te$_{3}$ \cite{shrestha2017extremely}, NbP \cite{shekhar2015extremely}, LaBi \cite{sun2016large}, etc.\ have attracted tremendous attention for studying their exotic topological properties \cite{chen2009experimental,PhysRevB.95.115140,fei2017edge}.\ In many reports,\ XMR effect is explained by compensation of electron and hole densities  \cite{ali2014large,PhysRevLett.113.216601,PhysRevB.93.235142} and non-trivial topological protection \cite{jiang2015signature,liang2015ultrahigh}.\ From two-band model,\ XMR effect is well established by  perfect electron-hole compensation \cite{ali2014large,PhysRevLett.113.216601,PhysRevB.93.235142}.\ Since, many XMR materials like LaSb, YSb, etc.\ are topologically trivial \cite{tafti2016resistivity,PhysRevLett.117.267201,Wadhwa_2019}, therefore the significance of topology in leading to XMR effect is yet to be established.\ Therefore,\ it would be very interesting to to find a XMR material having topologically trivial phase and a lack of perfect electron-hole compensation, so that non-trivial topological phase may be induced in it by enhancing the spin-orbit coupling (SOC) strength.\ 
In many recent reports,\ XMR effect is greatly pronounced in rare-earth monopnictide compounds \cite{sun2016large,tafti2016resistivity,ghimire2016magnetotransport,yu2017magnetoresistance,PhysRevB.97.081108,liang2018extreme,PhysRevB.97.085137,PhysRevB.99.245131} in which YSb have a lack of topological protection and perfect electron-hole compensation \cite{PhysRevLett.117.267201}.\ It is also known that chemical doping or alloying composition or applying pressure or strain can increase the strength of SOC \cite{sato2011unexpected,pal2014strain,doi:10.1002/adma.201605965,PhysRevB.96.081112,mondal2019emergence}, which may cause topological quantum phase transition (TQPT) in the material.\ But unlike chemical doping, external pressure is a strong tool to tune the electronic properties exempted from unwanted impurities arising from chemical doping.\ It inspired us to investigate the topological phase in YSb under hydrostatic pressure.\ In addition, we also studied electron to hole density ratio as a function of pressure, which may pave a path to correlate non-trivial band topology and XMR effect.\\

Vienna \textit{ab initio} simulation package (VASP) is used for electronic structure calculations within the framework of Density Functional Theory (DFT) \cite{PhysRevB.54.11169}.\ Exchange-correlation functions are included within the approximation of Perdew-Burke-Ernzehrof (PBE) under Generalized gradient approximation (GGA) and Heyd-Scuseria-Ernzerhof (HSE06) \cite{PhysRevLett.77.3865,heyd2003hybrid}.\ Electron-ion interactions are included within the formalism of projected augmented wave (PAW) \cite{kresse1999ultrasoft}.\ Theoretical calculations are performed with a cut-off energy of 300 eV.\ Forces on the each atom in the system are relaxed up to 0.001 eV.{\AA}$^{-1}$.\ We have used k-mesh of size 20$\times$20$\times$20 within Monkhorst-Pack formalism for sampling the Brillouin zone (BZ) under GGA.\ The variable cell relaxation method is used to simulate the system under different applied pressures.\ Spin-orbit coupling (SOC) is incorporated to determine the topological nature in the band structures.\ Electron to hole density ratio is calculated by plotting the Fermi surface at different applied pressures with a fine k-grid of 200$\times$200$\times$200 via constructing maximally localized wannier functions (MLWF) \cite{mostofi2014updated}. 

\section{\label{sec:level1}Results and Discussion}

YSb is found to have rocksalt crystal structure at ambient pressure with a space group of \textit{Fm$\bar{3}$m} (225) in which `Sb' atom is present at the origin (0, 0, 0), and `Y' atom is present at (0.5, 0.5, 0.5) \cite{ghimire2016magnetotransport}.\ The optimized lattice constant of YSb is found to be 6.201 \AA \ consistent with the experimental value of 6.163 \AA \cite{yu2017magnetoresistance,PhysRevB.97.085137}.\\

\begin{figure}[htb!]
 \centering
 \includegraphics[width=9cm]{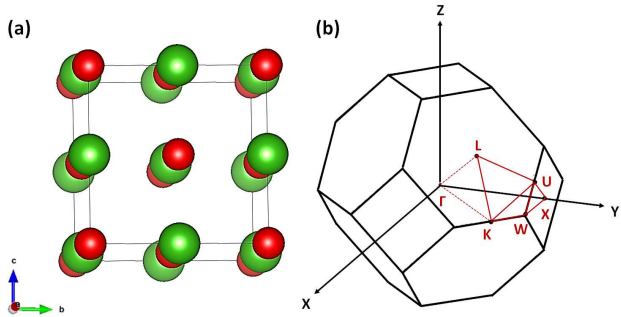}
 \caption{(color online) (a) Conventional unit cell of YSb.\ Red spheres denote `Sb' atoms and green spheres denote `Y' atoms. (b) First Brillouin zone of rocksalt crystal structure.}
 \label{figureone}
\end{figure}

\begin{figure*}[htb!]
 \includegraphics[width=14cm]{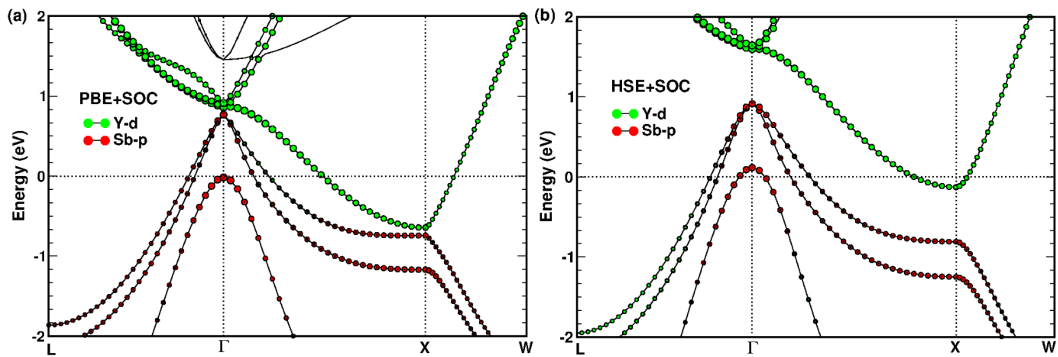}
 \centering
 \caption{\label{figuretwo}(color online) Band structures of YSb at ambient pressure calculated using exchange correlation functions of (a) PBE and (b) HSE including spin-orbit coupling.\ The \textit{p}-orbitals of `Sb' atom are denoted by red circles and \textit{d}-orbitals of `Y' atom are denoted by green circles.}
\end{figure*}

In order to investigate the non-trivial topological phase under pressure, first we calculated the band structure of YSb at ambient pressure using PBE and HSE functionals including the SOC effect as depicted in Figure \ref{figuretwo}.\\ 

It is found that valence band and conduction band of YSb are crossing the Fermi level for both the functionals, and are semi-metallic in nature \cite{ghimire2016magnetotransport,yu2017magnetoresistance}.\ It is observed that \textit{p}-orbitals of the `Sb' atom (represented by red circles) and \textit{d}-orbitals (t$_{2g}$) of `Y' (represented by green circles) have major contribution to the valence band and conduction band, respectively, near the Fermi level in both the band structures calculated using PBE as well as HSE functionals.\ Further, YSb shows a lack of band inversion for both the functionals,\ indicating that it is topologically trivial at ambient pressure, in agreement with the other reports \cite{yu2017magnetoresistance}.\\

\begin{figure*}[htb!]
 \centering
 \includegraphics[width =14cm]{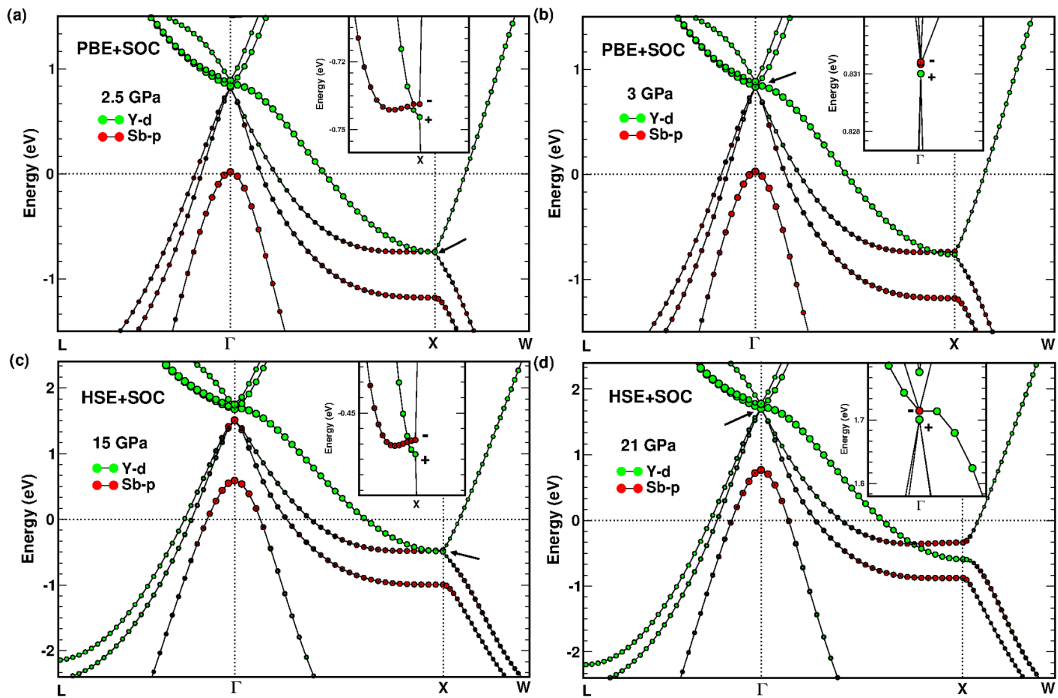}
 \caption{\label{figurethree}(color online) Band structures for YSb including spin-orbit coupling using PBE functionals at a pressure of (a) 2.5 GPa and (b) 3 GPa, and using HSE functionals at a pressure of (c) 15 GPa and (d) 21 GPa.\ The \textit{p}-orbitals of `Sb' atom are denoted by red circles and \textit{d}-orbitals of `Y' atom are denoted by green circles.}
\end{figure*}

On increasing the pressure in YSb, there is an increase in the overlap between their valence band and conduction band, and a band inversion between \textit{p}-orbitals of `Sb' atom and \textit{d}-orbitals (t$_{2g}$) of `Y' atom is observed on X point at 2.5 GPa using PBE functionals, while at 15 GPa using HSE functionals [Figure \ref{figurethree}(a) and (c)].\ On further increase in pressure, we found two band inversions between \textit{p}-orbitals of `Sb' atom and \textit{d}-orbitals (t$_{2g}$) of `Y' atom at $\Gamma$ and X points at 3 GPa using PBE functionals, while at 21 GPa using HSE functionals [Figure \ref{figurethree}(b) and (d)].\\

We have also calculated the Z$_{2}$ topological invariant for YSb to confirm the non-trivial topological phase at different applied pressures.\ In three dimensions,\ the topological insulator has four Z$_{2}$ invariants ($\nu_{0}$; $\nu_{1}$ $\nu_{2}$ $\nu_{3}$),\ and value of the first Z$_{2}$ topological invariant $\nu_{0}$ = 1 implies a strong topological phase, while $\nu_{0}$ = 0 implies either a weak TI phase or topologically trivial \cite{PhysRevLett.98.106803}.\ The value of the first Z$_{2}$ invariant $\nu_{0}$ for a system preserving  space-inversion and time-reversal symmetries is calculated by finding the parities of all the filled states at all time-reversal invariant momenta (TRIM) points \cite{PhysRevB.76.045302}.\ For a centro-symmetric system having time-reversal symmetry, there exists eight TRIM points in three-dimensions \cite{PhysRevB.76.045302}.\ The detailed parities for YSb at ambient pressure and its corresponding pressures of topological phase transitions using PBE and HSE functionals are shown in Table \ref{one}.\\


\begin{table*}[]
\caption{Detailed parities of YSb for all eight TRIM points in the BZ (a) at ambient pressure using both PBE and HSE functionals, (b) using PBE functionals at a pressure of (i) 2.5 GPa and (ii) 3 GPa, and (c) using HSE functionals at a pressure of (i) 15 GPa and (ii) 21 GPa. }
\label{one}
\begin{tabular}{|c|c|c|c|c|c|c|c|c|c|c|c|c|c|c|c|}
\hline
\multirow{3}{*}{\textbf{\begin{tabular}[c]{@{}c@{}}Band\\ No.\end{tabular}}} & \multicolumn{3}{c|}{\multirow{2}{*}{\textbf{\begin{tabular}[c]{@{}c@{}}(a) Ambient \\pressure\end{tabular}}}} & \multicolumn{6}{c|}{\textbf{(b) PBE}} & \multicolumn{6}{c|}{\textbf{(c) HSE}} \\ \cline{5-16} 
 & \multicolumn{3}{c|}{} & \multicolumn{3}{c|}{\textbf{(i) 2.5 GPa}} & \multicolumn{3}{c|}{\textbf{(ii) 3 GPa}} & \multicolumn{3}{c|}{\textbf{(i) 15 GPa}} & \multicolumn{3}{c|}{\textbf{(ii) 21 GPa}} \\ \cline{2-16} 
 & \textbf{4L} & \textbf{$\Gamma$} & \textbf{3X} & \textbf{4L} & \textbf{$\Gamma$} & \textbf{3X} & \textbf{4L} & \textbf{$\Gamma$} & \textbf{3X} & \textbf{4L} & \textbf{$\Gamma$} & \textbf{3X} & \textbf{4L} & \textbf{$\Gamma$} & \textbf{3X} \\ \hline
\textbf{1} & - & + & + & - & + & + & - & + & + & \textbf{-} & \textbf{+} & \textbf{+} & \textbf{-} & \textbf{+} & \textbf{+} \\ \hline
\textbf{3} & + & - & - & + & - & - & + & - & - & \textbf{+} & \textbf{-} & \textbf{-} & \textbf{+} & \textbf{-} & \textbf{-} \\ \hline
\textbf{5} & + & - & - & + & - & - & + & - & - & \textbf{+} & \textbf{-} & \textbf{-} & \textbf{+} & \textbf{-} & \textbf{-} \\ \hline
\textbf{7} & + & - & - & + & - & - & + & - & - & \textbf{+} & \textbf{-} & \textbf{-} & \textbf{+} & \textbf{-} & \textbf{-} \\ \hline
\textbf{9} & + & + & + & + & + & + & + & + & + & \textbf{+} & \textbf{+} & \textbf{+} & \textbf{+} & \textbf{+} & \textbf{+} \\ \hline
\textbf{11} & - & - & - & - & - & - & - & - & - & \textbf{-} & \textbf{-} & \textbf{-} & \textbf{-} & \textbf{-} & \textbf{-} \\ \hline
\textbf{13} & - & - & - & - & - & - & - & - & - & \textbf{-} & \textbf{-} & \textbf{-} & \textbf{-} & \textbf{-} & \textbf{-} \\ \hline
\textbf{15} & - & - & - & - & - & + & - & + & + & \textbf{-} & \textbf{-} & \textbf{+} & \textbf{-} & \textbf{+} & \textbf{+} \\ \hline
\textbf{Total} & + & + & + & + & + & - & + & - & - & \textbf{+} & \textbf{+} & \textbf{-} & \textbf{+} & \textbf{-} & \textbf{-} \\ \hline
\textbf{$\nu_{0}$} & \multicolumn{3}{c|}{\textbf{0}} & \multicolumn{3}{c|}{\textbf{1}} & \multicolumn{3}{c|}{\textbf{0}} & \multicolumn{3}{c|}{\textbf{1}} & \multicolumn{3}{c|}{\textbf{0}} \\ \hline
\end{tabular}
\end{table*}

\begin{figure}
 \centering
 \includegraphics[width=9cm]{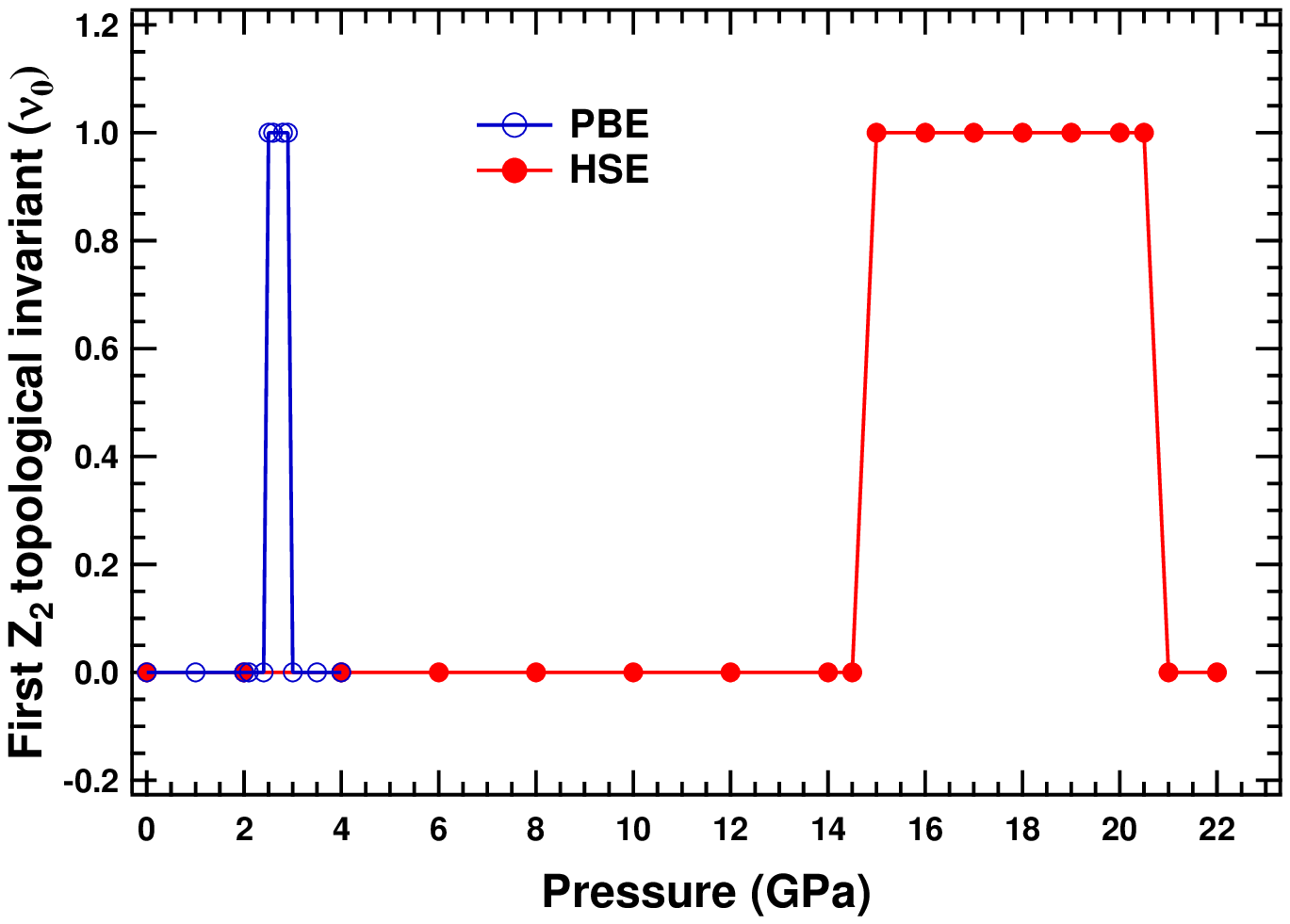}
 \caption{\label{figurefour}(color online) First Z$_{2}$ topological invariant ($\nu_{0}$) vs pressure (GPa) for YSb using PBE (blue) and HSE (red) functionals.}
\end{figure}

At ambient pressure, band structure of YSb shows no band inversion at any TRIM point using both the functionals, and its first Z$_{2}$ topological invariant ($\nu_{0}$) comes out to be 0, according to Kane and Mele criterion \cite{PhysRevLett.98.106803},\ indicating that it is topologically trivial.\ But at 2.5 GPa using PBE and 15 GPa using HSE functionals, we observed a band inversion behavior on X point, and $\nu_{0}$ changes to 1.\ With further increase in pressure to 3 GPa using PBE and 21 GPa using HSE, we observed two band inversions, one at $\Gamma$, other at X point, and $\nu_{0}$ again becomes 0, which indicates either the weak TI phase or topologically trivial phase of the system.\ To determine the nature of topological phase with even no. of band inversions, we need to calculate other three topological indices ($\nu_{1}$ $\nu_{2}$ $\nu_{3}$). Since, three X points and four L points in the first BZ have the same parity at 3 GPa (PBE)/ 21 GPa (HSE) in YSb, therefore the value of other three topological invariants ($\nu_{1}$ $\nu_{2}$ $\nu_{3}$) turns out to be (0 0 0).\ It indicates that YSb is topologically trivial at 3 GPa (PBE)/ 21 GPa (HSE).\ Thus, YSb shows reentrant topological phase under hydrostatic pressure.\ The plot of first Z$_{2}$ topological invariant vs pressure for YSb using PBE and HSE functionals is depicted in Figure \ref{figurefour}.\\
  
 The above results demonstrate that multiple topological phases can be realized in the same material on increasing pressure, which can be verified experimentally by performing Berry phase measurements.\ The observation of reentrant behavior in topological phase under hydrostatic pressure in XMR material YSb provides another platform to study it with electron-hole compensation, which may help in determining the correlation between topology and XMR effect.\\

\begin{figure}[htb!]
 \centering
 \includegraphics[width=9cm]{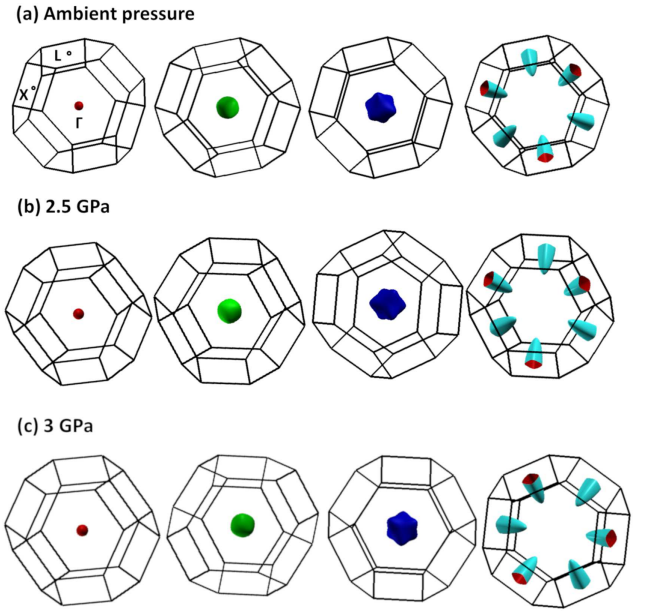}
 \caption{\label{figurefive}(color online) Fermi surface of YSb calculated using PBE functionals with spin-orbit coupling at (a) ambient pressure, (b) 2.5 GPa, and (c) 3 GPa.}
\end{figure}

To investigate electron to hole density ratio ($n_{e}/n_{h}$) as a function of hydrostatic pressure, Fermi surface needs to be plotted at different applied pressures.\ For this,\ first we calculated the Fermi surface of YSb at ambient pressure using both GGA and HSE functionals including the SOC effect.\ For PBE functionals, a k-mesh of 20$\times$20$\times$20 is used to construct the MLWF, while 10$\times$10$\times$10 is used for HSE functionals.\ The $n_{e}/n_{h}$ ratio for YSb at ambient pressure using PBE and HSE functionals comes out to be 0.947 and 0.819, respectively, where the value calculated from HSE functionals is well below the experimental value of 0.95 \cite{PhysRevB.96.075159}.\ It indicates that k-mesh of 10$\times$10$\times$10 using HSE functaionals is not sufficient to calculate the accurate Fermi surface.\ As HSE calculations are computationally very expensive for dense k-mesh and $n_{e}/n_{h}$ ratio calculated using GGA agrees well with the experiments, so we calculated $n_{e}/n_{h}$ ratio using PBE functionals to study electron-hole compensation as a function of pressure.\ Fermi surface of YSb calculated using PBE functionals including the SOC effect at ambient pressure, 2.5 GPa, and 3 GPa are shown in Figure \ref{figurefive}.\\

\setlength{\tabcolsep}{1em}

It can be seen from the band structures of YSb at ambient pressure and their corresponding pressures of topological phases [Figure \ref{figuretwo}(a), \ref{figurethree}(a), and \ref{figurethree} (b)] that three of the valence bands are crossing the Fermi level at $\Gamma$ point, thereby leading to three hole pockets at $\Gamma$ point.\ Meanwhile, only one of the conduction bands are crossing the Fermi level at X point in both the band structures of YSb, thereby forming only one electron pocket at X point [Figure \ref{figurefive}].\ By calculating the electron and hole pockets volume, we computed the density of electrons ($n_{e}$) and holes ($n_{h}$), and their ratio ($n_{e}/n_{h}$) from 0 to 4 GPa for YSb, and is provided in Table \ref{two}.\

\setlength{\tabcolsep}{1em}
\begin{table}[ht]
\centering
\caption{Density of electrons and holes (10$^{20}$ cm$^{-3}$) calculated using PBE functionals with spin-orbit coupling at different applied pressures, and their ratios for YSb.}
\label{two} 
\begin{tabular}{@{}cccc@{}}
\toprule
\textbf{\begin{tabular}[c]{@{}c@{}}Pressure \\ (GPa)\end{tabular}} & \textbf{\textit{n$_{e}$}} & \textbf{\textit{n$_{h}$}} & \textbf{\textit{n}$_{e}$/\textit{n}$_{h}$} \\ \midrule

0                       & 3.130 & 3.255 & 0.947        \\
1                       &  3.276                                               &                        3.415                        &   0.959           \\
2                       & 3.470 & 3.567 & 0.973        \\
2.5                     &  3.555      & 3.627      & 0.980   \\
3                       &       3.655                                         &                              3.699                  &  0.988      \\
4                       & 3.827 & 3.82 & 1.001        \\ \bottomrule
\end{tabular}
\end{table}





It is observed that the values of electron density ($n_{e}$), hole density ($n_{h}$), and their ratio calculated for YSb increases with pressure.\ From two-band model,\ magnetoresistance (MR) deduced for materials having both types of charge carriers is
\begin{equation}
MR = \frac{n_{e} \mu_{e}n_{h} \mu_{h}\left ( \mu _{e}+\mu _{h} \right )^{2}B^{2}}{\left (n_{e}\mu_{e}+n_{h} \mu_{h} \right )^{^{2}}+\left (n_{e}-n_{h} \right )^{2}\left (\mu_{e} \mu_{h} \right )^{2}B^{2}}\
\end{equation}
 where, $n_{e}$ and $\mu_{e}$ represents electron density and electron mobility, respectively, while $n_{h}$ and $\mu_{h}$ represents hole density and hole mobility, respectively; and B is the external magnetic field.\
 
For a given mobility of charge carriers,\ the magnetoresistance would be maximum for perfect electron-hole compensation \cite{PhysRevB.93.235142}.\ It is also reported that non-trivial topological protection forbids the electron backscattering at zero field, but opens backscattering path in finite magnetic field leading to the XMR effect \cite{jiang2015signature,liang2015ultrahigh}.\ From our calculations, we found that the ratio of $n_{e}/n_{h}$ for YSb increases with pressure, and becomes 0.98 at the pressure corresponding to the emergence of non-trivial phase at 2.5 GPa.\ It indicates that non-trivial topological phase appears without perfect electron-hole compensation.\ It is observed that the ratio of $n_{e}/n_{h}$ becomes 1.001 at 4 GPa, at which the system is topologically trivial.\ It shows that non-trivial topological phase may miss a maximal impact on XMR effect under hydrostatic pressure.\ Since, XMR depends upon the carrier density as well as mobility, and mobility also changes with pressure, therefore the evolution of mobility as a function of pressure is also needed to be explored in order to determine the exact correlation between topolgy and XMR effect, which can only be done via experiments.\

\section{\label{sec:level1}Conclusion}
It is concluded that XMR material YSb undergoes reentrant topological quantum phase transition under hydrostatic pressure.\ From the detailed Fermi surface calculations, it is found that the ratio of $n_{e}/n_{h}$ increases with the pressure, but perfect electron-hole compensation is absent in non-trivial topological phase.\ It indicates that non-trivial topological phase may appear without a maximal effect on the magnetoresistance under hydrostatic pressure, and need further experimental investigations to understand the exact relationship between non-trivial band topology and XMR effect.\ 

\section*{ACKNOWLEDGEMENTS}
The authors acknowledge IIT Ropar for providing High performance computing (HPC) facility.

\section*{References}
 \bibliographystyle{aipnum4-1}
 \bibliography{manuscript}

\end{document}